\documentclass[epj]{webofc}
\usepackage[varg]{txfonts}   % Web of Conferences font
\wocname{EPJ Web of Conferences}
\woctitle{New Frontiers in Physics 2014}
\newcommand{\be}{\begin{equation}}
\newcommand{\ee}{\end{equation}}
\newcommand{\bea}{\begin{eqnarray}}
\newcommand{\eea}{\end{eqnarray}}

\begin{document}
\selectlanguage{english}
\title{Gravitino Condensates in the Early Universe and Inflation}
%
% subtitle is optional

\author{Nick E. Mavromatos\inst{1,2}\fnsep\thanks{\email{Nikolaos.Mavromatos@kcl.ac.uk}}   
        %\and
        %Third author\inst{3}
        % etc.
}

\institute{King's College London, Department of Physics, Theoretical Particle Physics and Cosmology Group, Strand London WC2R 2LS, UK
\and
       CERN, Physics Department-Theory Division, CH 1211 Geneve 23, Switzerland.    
          }

\abstract{%
We review work on the formation of gravitino condensates via the super-Higgs effect in the early Universe. This is a scenario for both inflating the early universe and breaking local supersymmetry (supergravity), entirely independent of any coupling to external matter. 
The goldstino mode associated with the breaking of (global) supersymmetry is ``eaten'' by the gravitino field, which becomes massive (via its own vacuum condensation) and breaks the local supersymmetry (supergravity) dynamically. The most natural association of gravitino condensates with inflation proceeds in an indirect way, via a Starobinsky-inflation-type phase. The higher-order curvature corrections of the (quantum) effective action of gravitino condensates induced by integrating out massive gravitino degrees of freedom in a curved space-time background, in the broken-supergravity phase, are responsible for inducing a scalar mode which inflates the Universe. The scenario is in agreement with Planck data phenomenology in a natural and phenomenologically-relevant range of parameters, namely Grand-Unified-Theory values for the supersymmetry breaking energy scale and dynamically-induced gravitino mass.}
\maketitle

\section{Introduction and Summary} 

The inflationary paradigm is at present a successful one, offering an elegant solution to the so-called horizon and flatness problems of the standard Big Bang cosmology, whilst simultaneously seeding both the large-scale structure of the universe and temperature anisotropies of the CMB via quantum fluctuations occurring during the inflationary epoch.
The precise microscopic mechanism of inflation is however unknown at present.

The data favour, or - from a rather more conservative viewpoint - are in agreement with, a scalar field or fields with canonical kinetic terms slowly rolling down an almost flat potential in the context of Einstein gravity, generating in the process 50 - 60 e-folds of inflation, along with adiabatic, nearly scale invariant primordial density perturbations \cite{Planck,encyclo}.
From the best fit value of the running spectral index $n_s\sim0.96$ for the gravitational perturbations in the slow-roll parametrisation, found by Planck \cite{Planck},  and the usual relations among the slow-roll inflationary parameters~\cite{encyclo}
\begin{align}\label{spectral}
	n_s=1-6\epsilon+2\eta\,, \quad 
	r=16\epsilon\,,
\end{align}
we then find $r\lesssim0.11$ given the non-observation of primordial gravitational wave-like (transverse and traceless) perturbations by Planck or WMAP collaborations (that is the absence of B-mode polarisations). This observational fact implies that the energy scale $E_I$ of inflation is much smaller than 
the Planck scale $m_P$, lying in the ballpark of the Grand Unified Theory (GUT)  scale~\cite{Planck,encyclo}
$E_I \; = \; \Big(3\, H^2_I M_{\rm Pl}^2 \Big)^{1/4} \simeq 2.1 \times 10^{16} \times \Big(\frac{r}{0.20}\Big)^{1/4} \, {\rm GeV}$,
with $M_{\rm Pl} = 2.4 \times 10^{18}~{\rm GeV}$  the reduced Planck mass, $H_I$ the Hubble scale during inflation and $r$ the tensor-to-scalar perturbation ratio~\cite{encyclo}.The upper bound on $r < 0.11$ placed by the Planck Collaboration~\cite{Planck} implies 
\begin{equation}\label{upperH}
H_I  = 1.05 \, \Big(\frac{r}{0.20} \Big)^{1/2}\times 10^{14}~{\rm GeV} \, \leq    0.78 \times 10^{14}~{\rm GeV}~,
\end{equation}
that is an upper bound four order of magnitudes smaller than the reduced Planck mass.

An important issue at present is the extent to which this inflationary process is tied to physics at the Grand Unification  (GUT) scale, and in particular, to a possible supersymmetric phase transition occurring in the early universe. 
Links of supersymmetry to inflation may be arguably expected from the fact that supersymmetry provides a rather natural reason~\cite{natural} for 
the smallness (compared to Planck scale) value of the inflationary Hubble scale (\ref{upperH}). 
If supersymmetry is realised in nature however, it is certainly broken. 

It is known that simple realisations of global supersymmetry (SUSY) breaking, such as in the Wess-Zumino model~\cite{croon}, can provide, 
when embedded in gravitational environments, slow-roll models for inflation consistent with the Planck data~\cite{Planck}.
Rigorous embeddings of global SUSY to local supersymmetry (SUGRA) have also been considered and explored in the literature over the years in connection with various scenarios for inflation~\cite{sugra_inflation_review}, such as hybrid~\cite{sugra_hybrid}, chaotic~\cite{sugra_chaotic}, no-scale SUGRA/Starobinsky-like~\cite{sugra_staro}. 
In the latter case inflation is linked to higher curvature terms in the gravitational action (such as $R^2$ terms), as in the original
Starobinsky model~\cite{staro}, and others~\cite{confsugra,sugrainfl}.
Such models have been compared against the recently available data, with the conclusion that Planck data~\cite{Planck} compatibility is straightforward.

However, In March 2014, the scientific community has been stunned by claims from the BICEP2 collaboration~\cite{BICEP2} on the measurement for the first time of B-mode polarization in the cosmic microwave background radiation, which was interpreted 
as evidence for gravitational waves at the time of the last scattering,
with a tensor-to-scalar ratio $r = 0.16^{+0.06}_{-0.05}$ after dust subtraction.
The BICEP2 data are consistent with a scalar spectral index $n_s \simeq 0.96 $ and no appreciable running,
in agreement with the Planck data~\cite{Planck}, but 
the Hubble parameter during slow-roll inflation $H_I$
indicated by the BICEP2 data is  larger than then upper limit (\ref{upperH}), imposed by Planck, that is one has now 
$H_I^{\rm Bicep2}  \sim 0.94 \times 10^{14}~{\rm GeV}$  for $r=0.16$.
If confirmed by subsequent experiments, such a large value of $r$ would
exclude Starobinsky-type inflationary potentials and mostly favour $\phi^2$ tachyonic models for inflation~\cite{encyclo}. However, there is currently an active debate as to 
whether the BICEP2 signal truly represents primordial gravitational waves, 
or is polluted by Galactic foregrounds and gravitationally-lensed E-modes~\cite{MS}. According to such works, 
the BICEP2 signal could be compatible with a cosmology with $r \ll 0.1$
if there is a dust polarization effect as large as presently allowed by Planck~\cite{Planck} and other data.
These remarks have recently been reinforced by data on the foreground dust in the BICEP2
region released by the Planck collaboration~\cite{Planckrec},
which point to significant foreground pollution that would affect the interpretation of the
BICEP2 B-mode polarization data. One therefore needs to wait for the results of the
planned joint analysis by the Planck and BICEP2 teams, before any definite conclusions are drawn on this important issue.
Until this  is resolved, it is advisable to keep an open mind about the possible range of $r$ values that
models of inflation could yield.
Therefore it seems premature to abandon Starobinsky-like models as potential candidates for realistic models of inflation compatible with the data. 

In this talk we shall present a rather minimal inflationary scenario which is associated indirectly with a Starobinsky type inflation. 
The approach is documented in a series of previous publications~\cite{emdyno,ahm,ahmstaro}, 
and is based on  the possibility of dynamically breaking SUGRA solely by means of exploiting the four-gravitino interactions that characterise (any) supergravity action, via the fermionic torsion parts of the spin connection. 
The primary example, where the calculations of the effective potential were detailed, was that of $\mathcal{N}=1$, $D=4$ simple SUGRA without matter~\cite{Freedman,Nieuwenhuizen}.  The dynamical breaking process may be concretely realised by means of a phase transition from the supersymmetric phase where the bilinear $\langle\overline\psi_\mu\psi^\mu\rangle$ representing the effective scalar degree of freedom has zero vacuum expectation value, to one where $\sigma \equiv \langle\overline\psi_\mu\psi^\mu\rangle\neq0$. 
The quantum excitations about this condensate vacuum are then identified with a gravitino condensate scalar field. 
Since this must be an energetically favourable process to occur, it then follows that the effective potential experienced by the gravitino condensate must be locally concave about the origin. 

The corresponding one-loop effective potential of the gravitino condensate scalar field, obtained after integrating out fermionic (gravitino) and bosonic (graviton) degrees of freedom therefore has the characteristic form of a Coleman-Weinberg double well potential, offering the possibility of hilltop-type inflation, with the condensate field playing the role of the inflaton~\cite{emdyno,ahm}, which would be the simplest scenario. However,  
in order to guarantee a slow-roll inflation one needs unnaturally large values of the gravitino-condensate wave function renormalisation.
This prompted us to discuss a second, rather indirect way, by means of which the gravitino condensate is associated with inflation~\cite{ahmstaro}. This is realised via 
the higher (in particular quadratic-order) curvature corrections of the (quantum) effective action of the gravitino condensate field, obtained after integration of graviton and gravitino degrees of freedom in the massive gravitino phase. These corrections induce a 
Starobinsky-type inflation~\cite{staro}, which occurs for quite natural values of the parameters of the ${\mathcal N}=1$ SUGRA model and its variants, as we shall review below.

The structure of the talk is as follows: In Section \ref{sec:effpot} we review the formalism and physical concepts underlying dynamical breaking of SUGRA and the associated super-Higgs effect, within the context of simple 
four dimensional ${\mathcal N} =1$ models, including superconformal extensions thereof (with broken conformal symmetry) that are necessitated for phenomenological reasons, as explained in the text. In Section \ref{sec:infl} we discuss the simplest possible scenario for hilltop inflation, where gravitino condensate fields near the origin of the effective potential play the role of the inflation field. Unfortunately, for (phenomenologically desirable) supersymemtry breaking scales that are near or below the GUT scale, the model is compatible with slow roll for very large (unnatural) values of the condensate wave function renormalisation. This prompts us to discuss in Section \ref{sec:star} alternative scenarios for inflation of Starobinsky type that may occur in the massive gravitino phase, near the non-trivial minimum of the effective potential. In such scenarios, which are compatible with the Planck results, the role of the inflaton field is played by the scalar mode that describes the effects of scalar-curvature-square terms that characterise the gravitational sector of the effective action in the broken SUGRA phase, after integrating out the massive gravitinos. Finally, conclusions and outlook are presented in section \ref{sec:concl}.

\section{Super-Higgs effect and dynamical breaking of ${\mathcal N}=1$ SUGRA  \label{sec:effpot}} 

Our starting point is the $\mathcal{N}=1$ $D=4$ (on-shell) action for `minimal' Poincar\'e supergravity in the second order formalism, following the conventions of ref.~\cite{Freedman} (with explicit factors of the (dimensionful) gravitational constant $\kappa^2 = 8\pi {\rm G} = 1/M_{\rm Pl}^2$, in units $\hbar=c=1$, where $M_{\rm Pl}$ the reduced Planck mass in four space-time dimensions):
 \begin{align}\label{sugraction}
	&S_{\rm{SG}}=\int d^4x \,e \left(\frac{1}{2\kappa^2}R\left(e\right)-\overline\psi_\mu\gamma^{\mu\nu\rho}D_\nu\psi_\rho+\mathcal{L}_{\rm torsion}\right),\\\nonumber
	&\kappa^2=8\pi G\,
	\quad\gamma^{\mu\nu\rho}=\frac{1}{2}\left\{\gamma^\mu,\gamma^{\nu\rho}\right\}\,,
	\quad \gamma^{\nu\rho}=\frac{1}{2}\left[\gamma^\nu,\gamma^\rho\right]\,,
\end{align}
where $R(e)$ and $D_\nu\psi_\rho\equiv\partial_\nu\psi_\rho+\frac{1}{4}\omega_{\nu ab}\left(e\right)\gamma^{ab}\psi_\rho$ are defined via the torsion-free connection and, given the gauge condition 
\begin{equation} \gamma\cdot\psi=0~,
\label{gcond}
\end{equation}
one can write  
\begin{align}\label{torsion}
	\mathcal{L}_{\rm torsion}=-\frac{1}{16}\left(\left(\overline\psi^\rho\gamma^\mu\psi^\nu\right)\left(\overline\psi_\rho\gamma_\mu\psi_\nu+2\overline\psi_\rho\gamma_\nu\psi_\mu\right)\right)\times2\kappa^2\,,
\end{align}
arising from the fermionic torsion parts of the spin connection~\footnote{We note in passing that such four-fermion interactions are characteristic of any Einstein-Cartan theory of fermions in curved space-time~\cite{mercuri}. In fact, in a standard spin-1/2 fermion-gravity theory, the torsion-induced four fermion interactions assume a repulsive axial (pseudovector)-current-current form $-\left({\overline \psi }\gamma^\mu \gamma^5 \psi \right)\left({\overline \psi} \gamma_\mu \gamma^5 \psi \right)$. 
As we demonstrate in the Appendix of the second article in ref.~\cite{ahm}, a corresponding repulsive axial-current-current term for the gravitino torsion terms can also be obtained by appropriately utilising Fierz identities in analogy with the Einstein-Cartan theory. One thus would naively conclude that dynamical breaking of supergravity may not be possible. However, in all such theories, the Fierz identities among the fermions, including  gravitinos, make the actual coefficient of  such terms ambiguous. Only the non-perturbative physics can 
settle the value of the four-fermion terms~\cite{Wetterich}, and hence dynamical breaking of symmetry via scalar gravitino condensates is a realistic possibility.}.

Extending the action off-shell requires the addition of auxiliary fields to balance the graviton and gravitino degrees of freedom. 
These fields however are non-propagating and may only contribute to topic at hand through the development of scalar vacuum expectation values, which would ultimately be resummed into the cosmological constant.
Making further use of the gauge condition (\ref{gcond}) in concert with the Fierz identities (as detailed in \cite{ahm}), we may write 
\begin{align}
		\mathcal{L}_{\rm torsion}
		=\lambda_{\rm S}\left(\overline\psi^\rho\psi_\rho\right)^2
		+\lambda_{\rm PS}\left(\overline\psi^\rho\gamma^5\psi_\rho\right)^2
		+\lambda_{\rm PV}\left(\overline\psi^\rho\gamma^5\gamma_\mu\psi_\rho\right)^2
\end{align}
where the couplings $\lambda_{\rm S}$, $\lambda_{\rm PS}$ and $\lambda_{\rm PV}$ express the freedom we have to rewrite each quadrilinear in terms of the others via Fierz transformation. 
This freedom in turn leads to a known ambiguity in the context of mean field theory \cite{Wetterich}, which we addressed in \cite{ahm}, where we refer the reader for details.

Following the original ideas of dynamical symmetry breaking by Nambu and Jona-Lasinio~\cite{NJL}, we wish to linearise these four-fermion interactions via suitable auxiliary fields, e.g.
\begin{align}
	\frac{1}{4}\left(\overline\psi^\rho\psi_\rho\right)^2\sim\sigma\left(\overline\psi^\rho\psi_\rho\right)-\sigma^2\,,
\end{align}	
where the equivalence (at the level of the action) follows as a consequence of the subsequent Euler-Lagrange equation for the auxiliary scalar $\sigma$.
Our task is then to look for a non-zero vacuum expectation value $\langle\sigma\rangle$ which would serve as an effective mass for the gravitino. 
To induce the super-Higgs effect~\cite{DeserZumino} we also couple in the Goldstino associated to global supersymmetry breaking via the addition of
\begin{align}\label{goldstino}
	\mathcal{L}_\lambda=f^2\det\left(\delta_{\mu\nu}+\frac{i}{2f^2}\overline\lambda\gamma_\mu\partial_\nu\lambda\right)\bigg|_{\gamma\cdot\psi=0}=f^2+\dots\,,
\end{align}	
where $\lambda$ is the Goldstino, $\sqrt{f}$ expresses the scale of global supersymmetry breaking, and \dots\, represents higher order terms which may be neglected in our weak-field expansion of the determinant.
It is worth emphasising at this point the universality of \eqref{goldstino}; any model containing a Goldstino may be related to $\mathcal{L}_\lambda$ via a non-linear transformation \cite{Komargodski}, and thus the generality of our approach is preserved. 

Upon the specific gauge choice (\ref{gcond}) for the gravitino field
and an appropriate redefinition, one may eliminate any presence of the Goldstino field from the final effective 
action describing the dynamical breaking of local supersymmetry, except the cosmological constant term $f^2$ in (\ref{goldstino}), which serves as a reminder of the pertinent scale of supersymmetry breaking.
The non-trivial energy scale this introduces, along with the disappearance (through field redefinitions) of the Goldstino field from the physical spectrum and the concomitant development of a gravitino mass, characterises the super-Higgs effect. 

We may then identify in the broken phase an effective action
\begin{align}\label{finalaction}
	S=\frac{1}{2\kappa^2}\int d^4x \,e \left(\left(R\left(e\right)-2\Lambda\right)-\overline\psi_\mu\gamma^{\mu\nu\rho}D_\nu\psi_\rho+ m_{\rm dyn} \, \left(\overline\psi_\mu \psi^\mu\right)\right),	
\end{align}
where $\Lambda$ is renormalised cosmological constant, to be contrasted with the (negative) tree level cosmological constant 
\begin{align}
	 \Lambda_0\equiv\kappa^2\left(\sigma ^2  -f^2\right)~,
\end{align}
and $m_{\rm dyn} \propto \langle \sigma \rangle $ is a dynamically generated gravitino mass, the origin of which will be explained presently.
It is worth stressing at this point that $\Lambda_0$ must be negative due to the incompatibility of supergravity with de Sitter vacua; if SUGRA is broken at tree level, then of course no further dynamical breaking may take place.

For phenomenological reasons which have been outlined in detail in refs.~\cite{emdyno,ahm}, and we shall discuss below, we adopt an extension of ${\mathcal N}=1$ SUGRA which incorporates local supersymmetry in the Jordan frame, enabled by an associated dilaton superfield~\cite{confsugra}. 
The scalar component $\varphi$ of the latter can be either a fundamental space-time scalar mode of the gravitational multiplet, i.e. the trace of the graviton (as happens, for instance, in supergravity models that appear in the low-energy limit of string theories), or a composite scalar field constructed out of matter multiplets.
In the latter case these could include the standard model fields and their superpartners that characterise the Next-to-Minimal Supersymmetric Standard Model, which can be consistently incorporated in such Jordan frame extensions of SUGRA~\cite{confsugra}. 
Upon appropriate breaking of conformal symmetry, induced by specific dilaton potentials (which we do not discuss here), one may assume that the dilaton field acquires a non-trivial vacuum expectation value $\langle \varphi \rangle \ne 0 $. 
One consequence of this is then that in the broken conformal symmetry phase, the resulting supergravity sector, upon passing (via appropriate field redefinitions) to the Einstein frame is described by an action of the form (\ref{sugraction}), but with the coupling of the gravitino four-fermion interaction terms being replaced by 
\begin{equation}\label{tildcoupl}
\tilde \kappa\equiv e^{-\langle\varphi\rangle}\kappa~,
\end{equation}
while the Einstein term in the action carries the standard gravitational coupling $1/2\kappa^2$. 

\begin{figure}[h!!]
		\centering
		\includegraphics[width=0.7\textwidth]{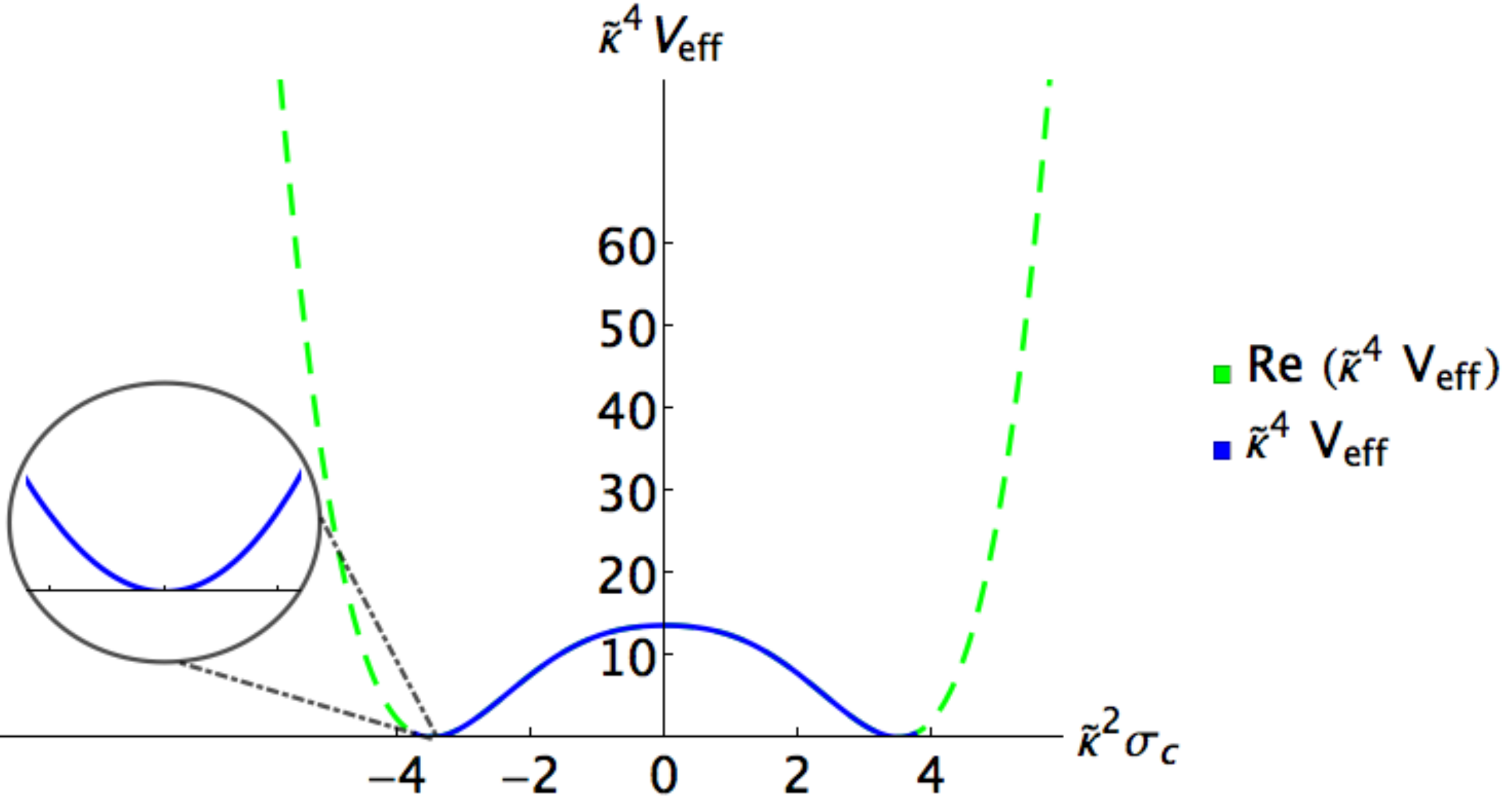} \vfill
		\includegraphics[width=0.4\textwidth]{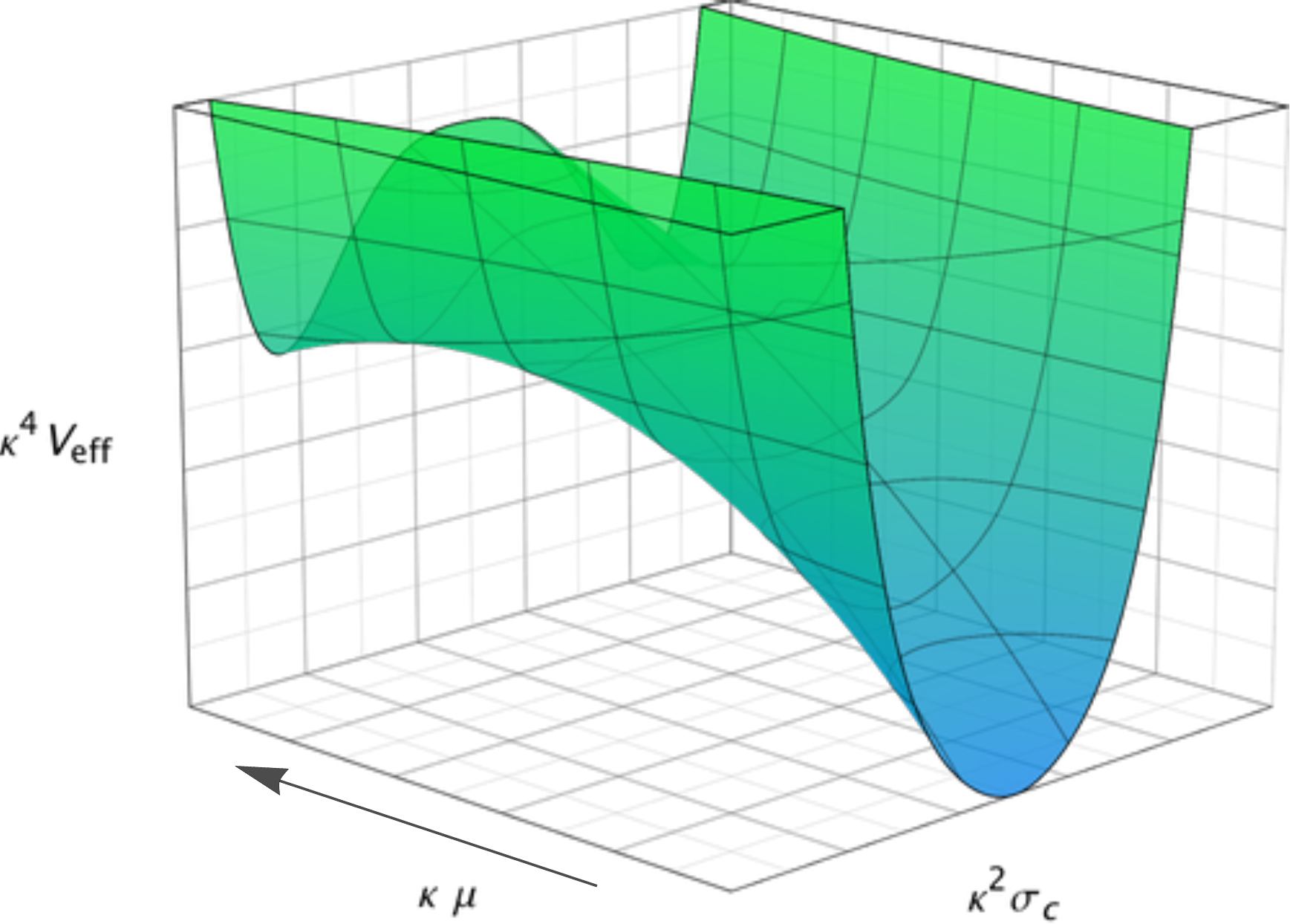}
		\includegraphics[width=0.4\textwidth]{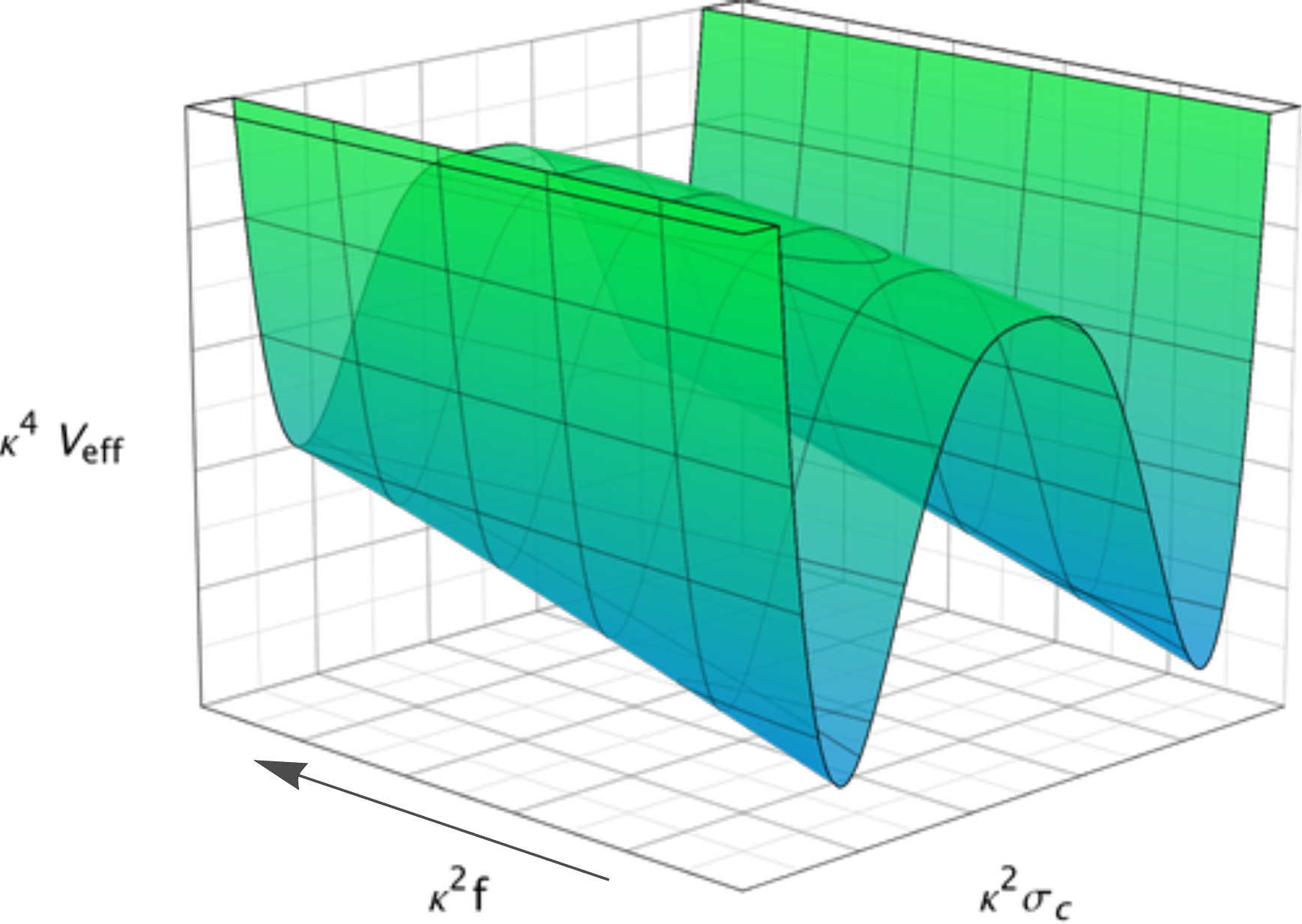}
		\caption{\emph{Upper panel}: The effective potential \eqref{effpotsugra}, expressed in units of the 
		coupling $\tilde \kappa$ (\ref{tildcoupl}). \emph{Lower panel}: As above, but showing schematically the effect of tuning the  RG scale $\mu$ and the supersymmetry breaking scale $f$, whilst holding, respectively, $f$ and $\mu$ fixed. The arrows in the respective axes correspond to the direction of increasing $\mu$ and $f$.}
\label{fig:effpotsugra}	
	\end{figure}

Expanding the graviton field about a de Sitter background~\cite{fradkin} (under the assumption that it is a solution of the one-loop effective equations) with renormalised cosmological constant $\Lambda > 0$, and integrating out both bosonic and fermionic quantum fluctuations to one loop yields the following effective potential for the gravitino condensate field $\sigma$ in the flat space-time limit $\Lambda \to  0$, as detailed in ref.~\cite{ahm}, 
\begin{align}\label{effpotsugra}
		V_{\text{eff}}=V_{B}^{(0)}+V_{B}^{(1)}+V_{F}^{(1)}
		=-\frac{\Lambda_0}{\kappa^2}+V_{B}^{(1)}+V_{F}^{(1)}~,
		\quad \Lambda_0\equiv\kappa^2\left(\sigma^2-f^2\right)~,
	\end{align}
where 	\begin{align}\label{boson}
		V_{B}^{(1)}=\frac{45 \kappa ^4}{512 \pi^2}\left(f^2-\sigma ^2\right)^2 \left(3-2 \ln \left(\frac{3 \kappa ^2 \left(f^2-\sigma ^2\right)}{2 \mu ^2}\right)\right)~, 
		\end{align} 
and 		\begin{align}\label{fermion}
		&V_{F}^{(1)} =\frac{{\tilde \kappa} ^4 \sigma^4}{30976 \pi ^2} \left(30578 \ln \left(\frac{{\tilde \kappa}^2 \sigma^2}{3 \mu^2}\right)-45867+29282 \ln \left(\frac{33}{2}\right)+1296 \ln\left(\frac{54}{11}\right)\right) \nonumber \\ 
& = \left(\frac{{\tilde \kappa}}{\kappa}\right)^4 \, \frac{\kappa^4 \sigma^4}{30976 \pi ^2} \left(30578 \ln \left(\left(\frac{{\tilde \kappa}}{\kappa}\right)^2\, \frac{\kappa^2 \sigma^2}{3 \mu ^2}\right)-45867+29282 \ln \left(\frac{33}{2}\right)+1296 \ln\left(\frac{54}{11}\right)\right)~,
	\end{align}
indicate the contributions to the effective potential from bosonic and fermionic fields respectively, and 	
$\mu$ is an inverse renormalisation group (RG) scale.
The effective potential (\ref{effpotsugra}) is depicted in fig.~\ref{fig:effpotsugra}. 

We may firstly note that as we flow from UV to IR (\emph{i.e}. in the direction of increasing $\mu$),  we obtain the correct double-well shape required for the super-Higgs effect, and secondly that tuning $f$ allows us to shift $V_{\text{eff}}$ and thus attain the correct vacuum structure (\emph{i.e.} non-trivial minima $\sigma_c$ such that $V_{\text{eff}}\left(\sigma_c\right)=0$).
Moreover,  the shape of the effective potential changes, as one varies the 
(renormalisation) scale $\mu$ from ultraviolet to infrared values (\emph{i.e.} flowing in the direction of increasing $\mu$), in such a way that the broken symmetry phase (double-well shaped potential) is reached in the IR. 
This indicates that the dynamical generation of a gravitino mass is actually an IR phenomenon, in accordance with the rather general features of dynamical mass in field theory. 

In the broken phase, the mass of the gravitino condensate is then given by 
\be
m_\sigma^2\equiv V''_{\text{eff}}(\sigma_c)~,
\ee
where $\sigma_c$ is the minimum of $V_{\text{eff}}$ and a prime denotes a functional derivative with respect to the gravitino-condensate field. As observed from (\ref{boson}), the bosonic contributions to the effective potential contain logarithmic terms which would contribute imaginary terms, leading to instabilities, unless 
\be\label{imagin}
\sigma_c^2<f^2~.
\ee
From \eqref{effpotsugra} it is straightforward to see that this condition is equivalent to the negativity of the tree-level cosmological constant $\Lambda_0$, which is entirely sensible; if $\Lambda_0>0$ then SUGRA is broken at tree level (given the incompatibility of supersymmetry with de Sitter vacua) and there can be no dynamical breaking.
As such, we must then tune $f$ for a given value of $\mu$ to find self consistent minima $\sigma_c$ satisfying (\ref{imagin}), thereby ensuring a real $V_{\text{eff}}$. In fact, here lies the importance of the super-Higgs effect, and thus of a non-zero positive $f^2 > \sigma_c^2 > 0$, in allowing dynamical breaking of local supersymmetry~\footnote{It should be mentioned at this point that in refs.~\cite{odintsov}, the importance of the super-Higgs effect was ignored, which led to the incorrect conclusion that imaginary parts exist necessarily in the one-loop effective potential (in the same class of gauges as the one considered in ref.~\cite{ahm} and here) and hence dynamical breaking of SUGRA was not possible. As we have seen above, such imaginary parts are absent when the condition (\ref{imagin}) is satisfied, and thus dynamical breaking of SUGRA occurs.}. 

As discussed in refs.~\cite{emdyno,ahm}, phenomenologically realistic situations, where one avoids transplanckian gravitino masses, for supersymmetry breaking scales $\sqrt{f} $ at most of order of the Grand Unification (GUT) scale $10^{15-16}$~GeV, as expected from arguments related to the stability of the electroweak vacuum, can occur only for 
large $\tilde \kappa$ couplings, typically of order $\tilde \kappa \sim \Big(10^{3}-10^{4} \Big)\, \kappa$. 
Given the relation \eqref{tildcoupl} this corresponds to dilaton vev of $\mathcal{O}\left(-10\right)$, where the negative sign may be familiar in the context of dilaton-influenced cosmological scenarios \cite{aben}.

If we consider for concreteness the case $\tilde\kappa=10^3 \kappa$, which is a value dictated by the inflationary phenomenology of the model~\cite{emdyno}, we may find solutions with a vanishing one-loop effective potential at the non-trivial minima corresponding to:
\begin{eqnarray}\label{solutions}
{\tilde \kappa}^2 \, \sigma_c \simeq 3.5~, \quad {\tilde \kappa}^2 \, f \simeq 3.7~, \quad {\tilde \kappa}\, \mu \simeq 4.0~, 
\end{eqnarray}
which leads to a global supersymmetry breaking scale 
	\begin{equation}\label{fscaleconf}
		\sqrt{f} \simeq 4.7\times10^{15}~{\rm GeV}~,
	\end{equation}
 and dynamical gravitino mass 
 	\begin{equation}\label{gravinoconf}
 		m_{\rm dyn}  
		\simeq 2.0\times10^{16}~{\rm GeV}~.
	\end{equation}
	
At the non-trivial minima we find $\tilde\kappa^4V_{F}^{(1)}\simeq-1.4$, $\tilde\kappa^4V_{B}^{(1)}\simeq5.9\times10^{-13}$, with tree-level cosmological constant $\tilde\kappa^2\Lambda_0 \simeq-1.4$.
We thus observe that fermion contributions to the effective potential are much stronger than the corresponding bosonic contributions for the cases of large couplings $\tilde \kappa \gg \kappa$. 
These values are phenomenologically realistic, thereby pointing towards the viability (from the point of view of producing realistic results of relevance to phenomenology) of the scenarios of dynamical breaking of local supersymmetry in conformal supergravity models~~\footnote{A comment concerning SUGRA models in the Jordan frame with such large values for their frame functions is in order here. In our approach, the dilaton $\varphi$ could be a genuine (dimensionless) dilation scalar field $\varphi = 2\phi$ arising in the gravitational multiplet of string theory, whose low-energy limit may be identified with some form of SUGRA action. In our normalization the string coupling would be $g_s \equiv e^\phi = {\tilde \kappa}^{-1/2}$. In 
such a case, a value of $\tilde \kappa = e^{-\langle \varphi \rangle}\, \kappa = {\mathcal O}(10^{3-4})$ would imply a large negative v.e.v. of the (four-dimensional) dilaton field of order $\langle \phi \rangle = -{\mathcal O}(5) < 0$, and thus a weak string coupling squared $g_s = {\mathcal O}(10^{-2})$, which may not be far from values attained in phenomenologically realistic string cosmologies~\cite{aben}. On the other hand, in the Jordan-frame SUGRA models of \cite{confsugra}, the frame function reads $\Phi \equiv e^{-\varphi} =  1 - \frac{1}{3}\Big(S{\overline S} + \sum_{u,d} \, H_i H^\dagger_i \Big) - \frac{1}{2} \chi \, \Big(-H_u^0 \, H_d^0 + H_u^+ \, H_d^- + {\rm h.c.} \Big)$,
in the notation of \cite{nmssm} for the various matter super fields of the next-to-minimal supersymmetric standard model that can be embedded in 
such supergravities. The quantity $\chi$ is a constant parameter. At energy scales much lower than GUT, it is expected that the various fields take on subplanckian values, in which case the frame function is almost one, and hence $\tilde \kappa \simeq \kappa $ for such models today. To ensure $\tilde \kappa \gg \kappa$,  and thus large values of the frame function, $\Phi \gg 1$, as required in our analysis, one needs to invoke trasnplanckian values for some of the fields, $H_{u,d}^0$, and large values of $\chi$, which may indeed characterize the inflationary phase of such theories. A similar situation occurs for the values of the higgs field (playing the role of the inflaton) in the non-supersymmetric Higgs inflation models~\cite{higgsinfl}.}.

On the other hand, in standard SUGRA scenarios, where $\tilde \kappa = \kappa$, one finds, as already mentioned, transplanckian values for the dynamically generated gravitino mass~\cite{ahm}:
$m_{\rm dyn} \simeq 2.0\times10^{19}~{\rm GeV}$, and a 
global supersymmetry breaking scale $\sqrt{f} \simeq 4.7\times10^{18}~{\rm GeV}$, far too high
to make phenomenological sense.

\section{Connection with Slow-Roll Hill-top Inflation \label{sec:infl}}

In order to discuss the possible connection with inflation, we need to calculate one more important ingredient; the wave-function renormalisation. In principle, this should be calculated in a curved de Sitter space-time, which characterises the (unbroken) phase of SUGRA, when the condensate field is near the trivial maximum of the effective potential (\ref{effpotsugra}). 
This is a complicated task. However, it turns out that, since, according to the data~\cite{Planck,encyclo}, the de Sitter phase Hubble parameter in phenomenologically relevant inflationary models is expected to be several orders of magnitude smaller than the Planck scale, $m_P$ (\ref{upperH}),
the space-time curvature during inflation is not too large, and thus a flat space-time estimate of the wave function renormalisation may suffice.

The effective Lagrangian describing the gravitino bound state, with a non-trivial wave-function renormalization, is 
\be
{\cal L}_{\rm eff}=\frac{Z\kappa^2}{2}\partial_\mu\sigma\partial^\mu\sigma-V_{\text{eff}}(\sigma)~,
\ee
where the rescaling $\sigma=\tilde\sigma/\kappa\sqrt{Z}$ leads to the canonically normalised Lagrangian
\be
\tilde{\cal L}_{\rm eff}=\frac{1}{2}\partial_\mu\tilde\sigma\partial^\mu\tilde\sigma-\tilde V_{\text{eff}}(\tilde\sigma)~,
\ee
and the coupling constants in the potential $\tilde V_{\text{eff}}$ are defined as
\be\label{Vn}
\tilde V_{\text{eff}}^{(n)}(0)\equiv\frac{V_{\text{eff}}^{(n)}(0)}{Z^{n/2}}~.
\ee
The latter normalisations ultimately yield the slow roll parameters 
\be\label{slowroll}
\epsilon=\frac{1}{Z}\frac{M_{\rm Pl}^2}{2}\left(\frac{V_{\text{eff}}'}{V_{\text{eff}}}\right)^2~,
~~~~\eta=\frac{1}{Z}M_{\rm Pl}^2\frac{V_{\text{eff}}''}{V_{\text{eff}}}~,
~~~~\xi=\frac{1}{Z^2}M_{\rm Pl}^4\frac{V_{\text{eff}}'V_{\text{eff}}'''}{V_{\text{eff}}^2}~.
\ee
That large values of $Z \gg 1$ are necessarily linked to slow-roll hilltop inflation in this case is to be expected from the fact that the  effective potential (\ref{effpotsugra}) can be approximated near the origin (\emph{i.e}. for 
small field values of the condensate $\tilde\sigma \to 0$) as:
\be
V_{\rm eff} \simeq f^2 - (Z\kappa^2)^{-1} {\tilde  \sigma}^2 ~, \quad \tilde \sigma \to 0~, 
\ee
for a canonically normalised condensate field $\tilde \sigma$. 
To ensure that the slow-roll parameter $|\eta| < 1$ (\ref{slowroll}), then,
we must have
\be\label{estimates} 
Z \gg \frac{M_{\rm Pl}^4}{f^2} ~.
\ee
Since the (observed) running spectra index is of order $\eta_s \simeq 0.96$~\cite{Planck}, we must further impose that $|\eta| < 10^{-2}$. 

As already mentioned, we assume that we can use the flat space-time results for the wave-function renormalisation obtained in \cite{ahm}, 
to obtain a correct order of magnitude estimate that is valid in the curved space-times during the inflationary period~\cite{emdyno}. We first note, that in the broken phase, with the phenomenologically acceptable values of the gravitino mass $m_{\rm dyn}$ and supersymmetry breaking scales $\sqrt{f}$
 the function $Z$ is of order one, which is consistent with the exit from the slow-roll inflationary phase.  
Near the origin of the potential (\ref{effpotsugra}), the wave-function renornmalization 
is~\cite{ahm}
\be\label{wfinfl}
Z \simeq-\frac{1}{2\pi^2}\ln\left(\omega_{\tilde \mu}^{2}\right)~, \quad \omega_{\tilde \mu} \equiv {\tilde \mu}/C_{\rm off}, \, 
\quad g\equiv\lambda_{\rm S} C_{\rm off}^2/2\pi^2~.
\ee  
where $\tilde \mu$ is a transmutation mass scale and $C_{\rm off}$ is a flat-space-time cutoff, which may be taken to be the Planck scale, for low energy theories. 
For appropriate values of $g$ this corresponds to the limit of large $Z \gg 1$ and small $\omega_{\tilde \mu}$, both of which are phenomenologically desirable~\footnote{Note that larger than one values of the wave-function renormalisation 
for the composite gravitino condensate fields do not contradict unitarity. A similar situation is encountered in composite Higgs symmetry breaking models in field theory~\cite{Bardeen}.}. 

\begin{figure}[h!!]
  \centering
    	\includegraphics[width=0.7\textwidth]{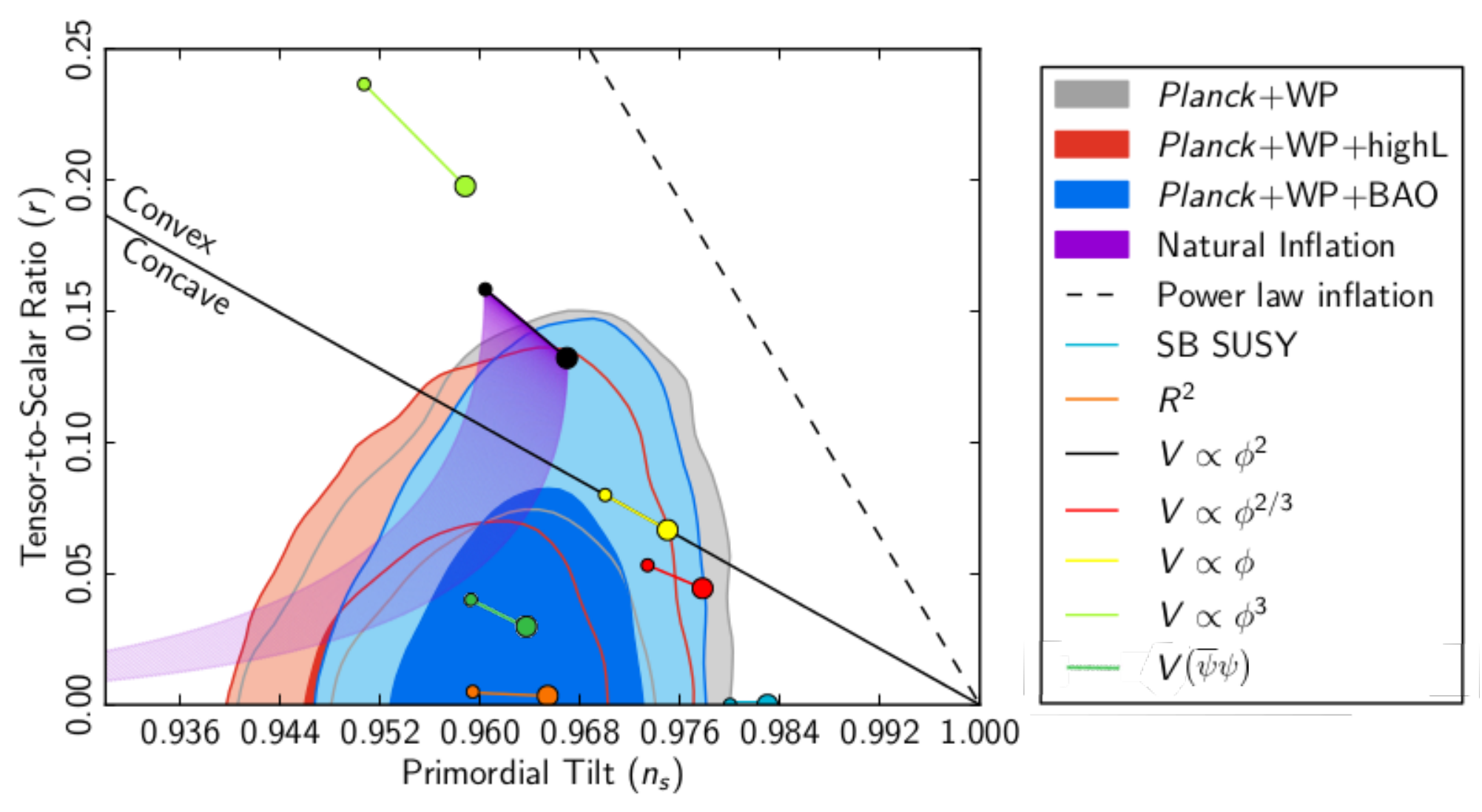}
    \caption{Planck data~\protect\cite{Planck} for $n_s$ and $r$ with the gravitino-condensate hill-top inflation indicated explicitly (dark green). The latter model leads to higher $r$ than Starobinsky-type $R^2$ inflation (orange), although requires a very high value of the gravitino-condensate wave-function renormalisation, of order larger than $O(10^{16})$.}
    \label{fig:planckexclude}
\end{figure}

Phenomenologically realistic models of broken SUSY have $\sqrt{f} < 10^{16}$~GeV = $10^{-2}~M_{\rm Pl}$ (\emph{cf}. (\ref{fscaleconf})), hence we must have $Z \gg 10^{10}$, implying very small, practically vanishing, transmutation mass scales. 
A typical case, compatible with the phenomenologically acceptable values
(\ref{fscaleconf}) and (\ref{gravinoconf}) is given in figure \ref{fig:planckexclude}, from which we observe that
agreement with Planck results is achieved for values of the wave function renormalisation of order
$Z \sim {\mathcal O }\left(10^{16}\right)$ 
for the phenomenologically relevant values of the couplings 
${\tilde \kappa }/\kappa \sim 10^3$. This corresponds to practically zero transmutation mass scales of $\tilde \mu \to 0$. It may be interesting to notice that increasing ${\tilde \kappa }/\kappa$ higher still has the effect of scaling $V_{\rm eff}$ whilst leaving the shape of the potential qualitatively unchanged, allowing smaller and smaller values of $\sqrt{f}$ and $m_{\rm dyn}$.
Whilst this decrease in $f$ tends to naturally increase the slow-roll parameter $\eta$, by virtue of \eqref{estimates} this scenario may still be rendered compatible with slow-roll inflation if $Z$ is scaled accordingly to counteract this. 
As such, Planck compatible inflation as demonstrated in figure \ref{fig:planckexclude} can be achieved for any value of ${\tilde \kappa }/\kappa$. 
The Planck-compatible result is $(0.959, 0.04)\le \{n_s, r\} \le (0.964, 0.03)$ for 50 and 60 e-folds, respectively, corresponding to $\sqrt{f} \sim 5 \, e^{\langle \varphi \rangle} \times 10^{18}$ GeV. This is the case for any value of the (negative) dilaton vev $\langle \varphi \rangle$, however, as already mentioned, for realistic supersymmetry breaking phenomenology one should really fix $\sqrt{f}$ around or below the GUT scale. 

One way to interpret this result is the following. 
Near the origin of the potential one is in the unbroken phase, and hence the gravitino condensate has not yet fully formed, or rather is beginning to form, corresponding to a very small value of the gravitino mass. 
This small value grows in actual time, until the condensate sits in the minimum of the potential after rolling downhill, at which point the gravitino mass is stabilised at phenomenologically acceptable value, \emph{e.g}. of order the GUT scale. 
The duration of the whole process is that of the slow-roll inflation period, and exit from this phase occurs near the non-trivial minimum of the potential (\ref{effpotsugra}).

One may object to the huge value of the wave function renormalisation (\ref{estimates}) during the slow-roll inflationary phase as unnatural~\footnote{One may be tempted to discuss, within the context of our minimal model, an alternative scenario, according to which global SUSY breaks at a transplanckian scale $\sqrt{f} \gg 1$ (in Planck units). In this case, the SUSY matter partners would completely decouple from the low-energy spectrum,
and hence there would be no experimental evidence for SUSY. On the other hand, local SUSY (SUGRA) would ensure inflation via the gravitino condensation mechanism described in this work, while the induced transplanckian dynamical mass for the gravitino, would remove any possibility of observing it as well.   From (\ref{estimates}) we can then conclude that slow-roll inflation could be achieved for natural values of the wave-function renormalisation $Z < {\mathcal O}(10)$, but in this case the stability of the electroweak vacuum would be delinked from any SUSY arguments. 
One could also try to relax the slow-roll assumption but this opens up a whole new game, where comparison with data may be complicated, 
and we do not consider it here.}. There are alternative scenarios of slow-roll inflation linked to this model which do not require such large $Z$, which we shall now come to discuss. 
These are associated with another type of inflation that may occur in the broken SUGRA phase, where, in contrast to the hill-top inflationary scenario discussed so far, the gravitino condensate field lies near its value that minimises the potential (\ref{effpotsugra}). 
In this scenario, the inflaton field is not the gravitino condensate, but it is linked to the scalar mode that parametrises a $R^2$-Starobinsky-like~\cite{staro} inflation that is associated with the effective gravitational action 
obtained after integrating out the massive gravitino-condensate degrees of freedom. 
This scenario 
was discussed in detail in ref.~\cite{ahmstaro}, and we now proceed to review it briefly.  

\section{Starobinsky-type inflation in the broken SUGRA phase \label{sec:star}}

Starobinsky inflation is a model for obtaining a de Sitter (inflationary) cosmological solution to gravitational equations arising from a (four space-time-dimensional) action that includes higher curvature terms. 
Specifically, an action of the type in which the quadratic curvature corrections consist only of scalar curvature terms~\cite{staro}
\begin{eqnarray}\label{staroaction}
{\mathcal S} = \frac{1}{2 \, \kappa^2 } \, \int d^4 x \sqrt{-g}\,  \left(R  + \beta  \, R^2 \right) ~,~ 
\beta = \frac{8\, \pi}{3\, {\mathcal M}^2 }~,
\end{eqnarray}
where $\kappa^2=8\pi G$, and ${\rm G}=1/m_P^2$ is Newton's (gravitational) constant in four space-time dimensions, with $m_P$ the Planck mass, and ${\mathcal M}$ is a constant of mass dimension one, characteristic of the model. 

The important feature of this model is that inflationary dynamics are driven purely by the gravitational sector, through the $R^2$ terms, 
and that the scale of inflation is linked to ${\mathcal M}$. From a microscopic point of view, the scalar curvature-squared terms in (\ref{staroaction}) are viewed as the result of quantum fluctuations (at one-loop level)  of conformal (massless or high energy) matter fields of various spins, which have been integrated out in the relevant path integral in a curved background space-time~\cite{loop}. The quantum mechanics of this model, proceeding by means of tunnelling of the Universe from a state of ``nothing'' to the inflationary phase of ref.~\cite{staro} has been discussed in detail in ref.~\cite{vilenkin}.
The above considerations necessitate truncation to one-loop quantum order and to curvature-square (four-derivative) terms, which 
implies that there must be a region of validity for curvature invariants such that $\mathcal{O}\big(R^2/m_p^4\big) \ll 1$. 
This is of course a condition satisfied in phenomenologically realistic scenarios of inflation~\cite{Planck,encyclo}, for which the inflationary Hubble scale $H_I $ satisfies (\ref{upperH}) (the reader should recall that $R = 12 H_I^2$ in the inflationary phase). 

\begin{figure}[h!!!]
\centering
		\includegraphics[width=0.5\textwidth]{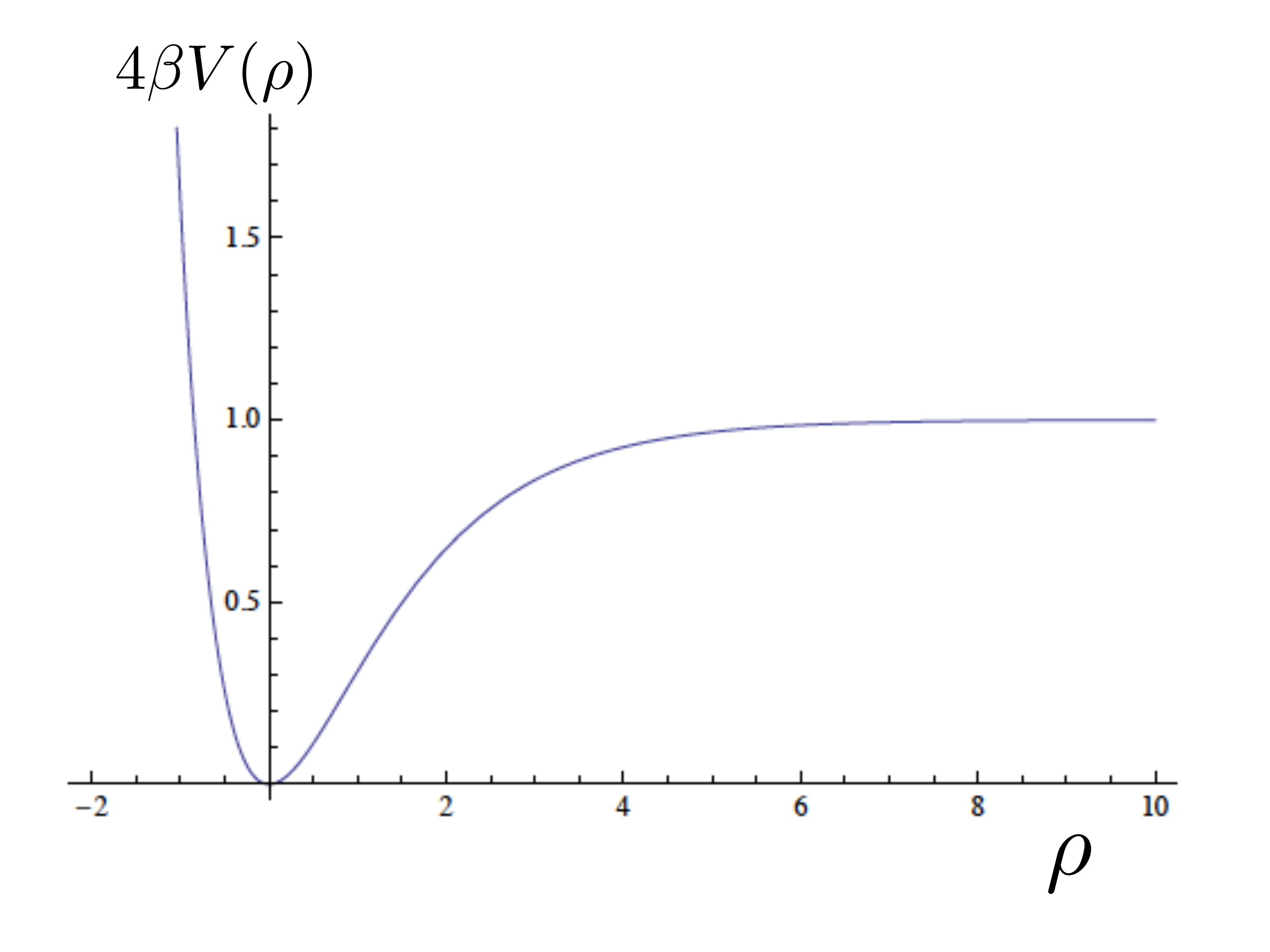}
		\caption{The effective potential (\ref{staropotent}) of the collective scalar field $\rho$ that describes the one-loop quantum fluctuations of matter fields, leading to the higher-order scalar curvature corrections in the Starobinski model for inflation (\ref{staroaction}). The potential is sufficiently flat to ensure slow-roll conditions for inflation are satisfied, in agreement with the Planck data, for appropriate values of the scale $1/\beta \propto {\mathcal M}^2$ (which sets the overall scale of inflation in the model).}
\label{fig:potstar}	
\end{figure}

Although the inflation in this model is not driven by rolling scalar fields, nevertheless the model (\ref{staroaction}) (and for that matter, any other model where the Einstein-Hilbert space-time Lagrangian density is replaced by an arbitrary function $f(R)$ of the scalar curvature) is conformally equivalent to that of an ordinary Einstein-gravity coupled to a scalar field with a potential that drives inflation~\cite{whitt}. 
To see this, one firstly linearises the $R^2$ terms in (\ref{staroaction}) by means of an auxiliary (Lagrange-multiplier) field $\tilde \alpha (x)$, before rescaling the metric by a conformal transformation and redefining the scalar field (so that the final theory acquires canonically-normalised Einstein and scalar-field terms):
\begin{eqnarray}\label{confmetric}
&&g_{\mu\nu} \rightarrow g^E_{\mu\nu} = \left(1 + 2 \, \beta \, {\tilde \alpha (x)} \right) \, g_{\mu\nu} ~, \quad
 \tilde \alpha \left(x\right) \to \rho (x) \equiv \sqrt{\frac{3}{2}} \, {\rm ln} \, \left(1 + 2\, \beta \, {\tilde \alpha \left(x\right)} \right)~.
\end{eqnarray}
These steps may be understood schematically via
\begin{align}\label{steps}
	\int d^4 x \sqrt{-g}\,  \left( R  + \beta  \, R^2 \right) 
%  	&\hookrightarrow\int d^4 x \sqrt{-g}\,  \left(  \left(1 + 2\, \beta \, \tilde \alpha \left(x\right) \right) \, R  -  \beta  \, {\tilde \alpha (x)}^2 \right)\\\nonumber
   \hookrightarrow\int d^4 x \sqrt{-g^E}\,  \left(R^E +  g^{E\, \mu\, \nu} \, \partial_\mu \, \rho \, \partial_\nu \, \rho - V\right(\rho\left) \right)~,
\end{align}
where the arrows have the meaning that the corresponding actions appear in the appropriate path integrals, 
with the potential $V(\rho)$ given by:
\begin{eqnarray}\label{staropotent}
 V(\rho ) = \frac{\left( 1 - e^{-\sqrt{\frac{2}{3}} \, \rho } \right)^2}{4\, \beta} \, 
  = \frac{3 {\mathcal M}^2 \, \Big( 1 - e^{-\sqrt{\frac{2}{3}} \, \rho } \Big)^2}{32\, \pi }  \,  ~.
\end{eqnarray}
The potential is plotted in fig.~\ref{fig:potstar}. 
We observe that it is sufficiently flat for large values of $\rho$ (compared to the Planck scale) to produce phenomenologically acceptable inflation, with the scalar field $\rho$ playing the role of the inflaton. 
In fact, the Starobinsky model fits the Planck data on inflation~\cite{Planck} well.

The agreement of the model of ref.~\cite{staro} with the Planck data has triggered an enormous interest in the current literature
in revisiting the model from various points of view, such as its connection with no-scale supergravity~\cite{sugra_staro} and (super)conformal versions of supergravity and related areas~\cite{sugra_chaotic,sugrainfl}. 
In the latter works however the Starobinsky scalar field is fundamental, arising from the appropriate scalar component of some chiral superfield that appears in the superpotentials of the model. 
Although of great value, illuminating a connection between supergravity models and inflationary physics, and especially for explaining the low-scale of inflation compared to the Planck scale, it can be argued that these works contradict the original spirit of the Starobinsky model (\ref{staroaction}). 
Therein, higher curvature corrections are viewed as arising from quantum fluctuations of matter fields in a curved space-time background, such that inflation is driven by the pure gravity sector in the absence of fundamental scalars. 

\begin{figure}[h!!!]
\centering
		\includegraphics[width=0.5\textwidth]{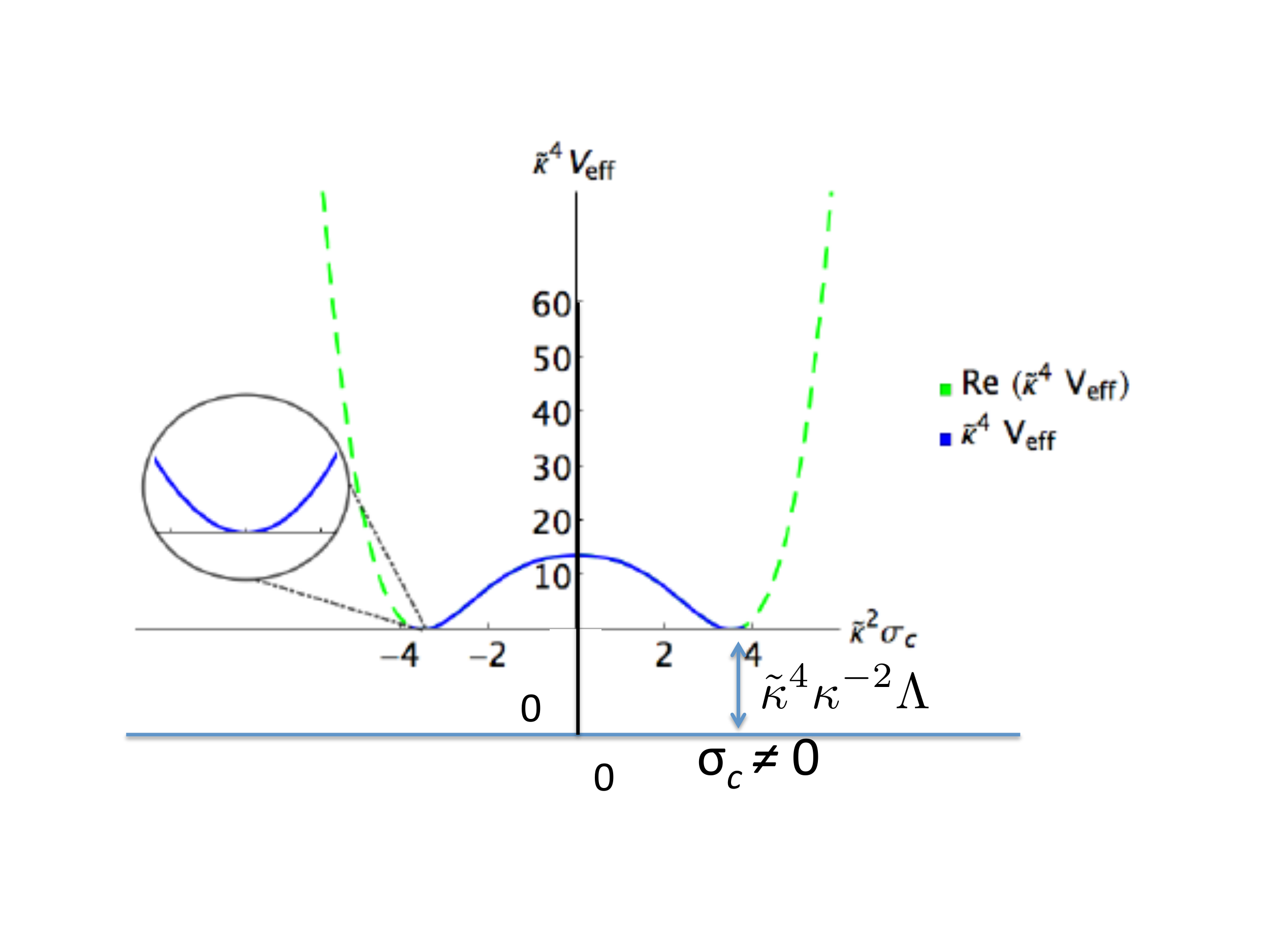}
		\caption{Generic shape of the one-loop effective potential for the gravitino condensate field $\sigma_c$ in dynamically broken (conformal) Supergravity models in the presence of a non-trivial de Sitter background with cosmological constant $\Lambda > 0$~\cite{ahmstaro}. The Starobinsky inflationary phase is associated with fluctuations of the condensate and gravitational field modes near the non-trivial minimum of the potential, where the condensate $\sigma_c \ne 0$, and the potential assumes the value $\Lambda > 0$, consistent with supersymmetry breaking. The dashed green lines denote ``forbidden'' areas of the condensate field values, violating the condition (\ref{imagin}), for which imaginary parts appear in the effective potential, thereby destabilising the broken symmetry phase.}
\label{fig:pot2}	
\end{figure}

In this section we consider an extension of the analysis of ref.~\cite{ahm} to the case where the de Sitter parameter $\Lambda$ is perturbatively small compared to $m_P^2$, but not zero, so that truncation of the series to order $\Lambda^2$ suffices. 
This is in the spirit of the original Starobinsky model~\cite{staro}, with the r\^ole of matter fulfilled by the now-massive gravitino field.
Specifically, we are interested in the behaviour of the effective potential near the non-trivial minimum, where $\sigma_c $ is a non-zero constant. 

It is important to notice at this point that, in contrast to the original Starobinsky model~\cite{staro}, where the crucial for inflation $R^2$ terms 
have been argued to arise from  the \emph{conformal anomaly} in the path integral of massless (conformal) matter in a de Sitter background, 
and thus their coefficient was arbitrary, in our scenario, such terms arise in the one-loop effective action of the gravitino condensate field, evaluated in a de Sitter background, after integrating out massive gravitino fields, whose mass was generated dynamically. The order of the de Sitter cosmological constant,
$\Lambda > 0$ that breaks supersymmetry, and 
the gravitino mass are all evaluated dynamically (self-consistently) in our approach from the minimization of the effective potential. Thus, the resulting $R^2$ coefficient, which determines the phenomenology of the inflationary phase,  is calculable~\cite{ahmstaro}. 
Moreover, in our analysis, unlike Starobinsky's original work, we will keep the contributions from both graviton (spin-two) and gravitino quantum fluctuations.
Specifically, we are interested in the behaviour of the effective potential near the non-trivial minimum, where $\sigma_c $ is a non-zero constant (\emph{cf} fig. \ref{fig:pot2}). 

The one-loop effective potential, obtained by integrating out gravitons and (massive) gravitino fields in the scalar channel (after appropriate euclideanisation), may be expressed as a power series in $\Lambda$: 
\begin{align}\label{effactionl2}
	\Gamma\simeq S_{\rm cl}-\frac{24\pi^2}{\Lambda^2 }\big(&\alpha^F_0+\alpha_0^B
	+ \left(\alpha^F_{1}+ \alpha^B_{1}\right)\Lambda
	+\left(\alpha^F_{2}+ \alpha^B_{2}\right)\Lambda^2+\dots\big)~,
\end{align} where $S_{\rm cl}$ denotes the classical action with tree-level cosmological constant $\Lambda_0$: 
	\begin{align}\label{l0def}
		-\frac{1}{2\kappa^2}\int d^4 x \sqrt{g}\left(\widehat{R}-2\Lambda_0\right), \quad
		\Lambda_0\equiv\kappa^2\left(\sigma^2-f^2\right)~,
	\end{align} 
with $\widehat R$ denoting the fixed $S^4$ background we expand around ($\widehat R=4\Lambda$, Volume = $24\pi^2/\Lambda^2$), and the $\alpha$'s indicate the bosonic and fermionic quantum corrections at each order in $\Lambda$.

The leading order term in $\Lambda$ is then the effective action found in \cite{ahm} in the limit $\Lambda\to0$, 
	\begin{align}
		\Gamma_{\Lambda\to0}\simeq-\frac{24\pi^2}{\Lambda^2}\left(-\frac{\Lambda_0}{\kappa^2}+\alpha_0^F+\alpha_0^B\right)
		\equiv\frac{24\pi^2}{\Lambda^2}\frac{\Lambda_1}{\kappa^2},
	\end{align}
and the remaining quantum corrections then, proportional to $\Lambda$ and $\Lambda^2$ may be identified respectively with Einstein-Hilbert $R$-type and Starobinsky $R^2$-type terms in an effective action (\ref{effactionl3}) of the form
\begin{align}\label{effactionl3}
\Gamma\simeq&-\frac{1}{2\kappa^2} \int d^4 x \sqrt{g} \left(\left(\widehat R-2\Lambda_1\right)  +\alpha_1 \, \widehat R+ \alpha_2 \, \widehat R^2\right)~,
\end{align}
where we have combined terms of order $\Lambda^2$ into curvature scalar square terms. For general backgrounds such terms 
would correspond to invariants of the form ${\widehat R}_{\mu\nu\rho\sigma} \, {\widehat R}^{\mu\nu\rho\sigma} $, ${\widehat R}_{\mu\nu} \, {\widehat R}^{\mu\nu}$ and ${\widehat R}^2$, which for a de Sitter background all combine to yield ${\widehat R}^2$ terms. 
The coefficients $\alpha_1$  and $\alpha_2$ absorb the non-polynomial (logarithmic) in $\Lambda$ contributions, so that we may then identify \eqref{effactionl3} with \eqref{effactionl2} via 
	\begin{align}\label{alpha}
		\alpha_1=\frac{\kappa^2}{2}\left(\alpha^F_1+\alpha^B_1\right)~,\quad
		\alpha_2=\frac{\kappa^2}{8}\left(\alpha^F_2+\alpha^B_2\right)~.
	\end{align}

To identify the conditions for phenomenologically acceptable Starobinsky inflation around the non-trivial minima of the broken SUGRA phase 
of our model, we impose first the \emph{cancellation} of the ``classical'' Einstein-Hilbert space term $\widehat R $ by the ``cosmological constant'' term $\Lambda_1$, \emph{i.e}. that 
$\widehat R = 4 \, \Lambda = 2\, \Lambda_1 $.
This condition should be understood as a necessary one characterising our background in order to produce phenomenologically-acceptable 
Starobinsky inflation in the broken SUGRA phase following the first inflationary stage, as discussed in ref.~\cite{emdyno}. 
This may naturally be understood as a generalisation of the relation $\widehat R=2\Lambda_1=0$, imposed in ref.~\cite{ahm} as a self-consistency condition for the dynamical generation of a gravitino mass. From thid it follows that the cosmological constant $\Lambda$ satisfies the four-dimensional Einstein equations in the non-trivial minimum, and in fact coincides with the value of the one-loop effective potential of the gravitino condensate at this minimum. As we discussed in \cite{ahm}, this non-vanishing positive value of the effective potential is consistent with the generic features of dynamical breaking of supersymmetry~\cite{witten}. In terms of the Starobinsky inflationary potential
(\ref{staropotent}), the value $\Lambda > 0$ corresponds to the approximately constant value of this potential in the high $\varphi$-field regime of fig.~\ref{fig:potstar}, where Starobinsky-type inflation takes place. Thus we may set 
$\Lambda \sim 3\, H_I^2 $, where $H_I$ the (approximately) constant Hubble scale during inflation, which is constrained by the current data to satisfy (\ref{upperH}).

The effective Newton's constant in  (\ref{effactionl3}) is then $\kappa_{\rm eff}^2=\kappa^2/\alpha_1$, and from this, we can express the effective Starobinsky scale (\ref{staroaction}) in terms of $\kappa_{\rm eff}$ as $\beta_{\rm eff} \equiv  \alpha_2/\alpha_1$.
This condition thus makes a direct link between the action (\ref{effactionl2}) with a Starobinsky type action (\ref{staroaction}).
Comparing with (\ref{staroaction}), we can then identify the Starobinsky inflationary scale in this case as
\begin{equation}\label{staroours}
{\mathcal M} = \sqrt{\frac{8 \pi}{3} \, \frac{\alpha_1}{\alpha_2} }~.
\end{equation}

We may then determine the coefficients $\alpha_1$ and $\alpha_2$ in order to evaluate the scale $1/\beta$ of the effective Starobinsky potential given in fig.~\ref{fig:potstar} in this case, and thus the scale of the second inflationary phase. 
To this end, we use the results of ref.~\cite{ahm}, derived via an asymptotic expansion as explained in the appendix therein, to obtain the following forms for the coefficients 		
	\begin{eqnarray}\label{aif}
\alpha^F_1&=& 0.067\, \tilde\kappa^2 \sigma_c ^2  -0.021\, \tilde\kappa^2 \sigma_c ^2 \, {\rm ln} \left(\frac{\Lambda}{\mu^2}\right) 
 + 0.073\, \tilde\kappa^2 \sigma_c ^2 \, {\rm ln} \left(\frac{\tilde\kappa^2\sigma_c^2}{\mu^2} \right)~, 
\nonumber \\
\alpha^F_{2}&=& 0.029 + 0.014\, {\rm ln} \left(\frac{\tilde\kappa^2\sigma_c^2}{\mu^2}\right)  -0.029\, {\rm ln} \left(\frac{\Lambda}{\mu^2}\right)~,
\end{eqnarray}
and 
	\begin{eqnarray}\label{aib}
\alpha^B_1&=& -0.083 \Lambda_0 + 0.018\, \Lambda_0 \, {\rm ln} \left(\frac{\Lambda }{3 \mu ^2}\right)  + 0.049\, \Lambda_0\,  {\rm ln} \left(-\frac{3 \Lambda_0}{\mu ^2}\right)~, 
\nonumber \\
\alpha^B_{2}&=& 0.020 +  0.021\, {\rm ln} \left(\frac{\Lambda }{3 \mu ^2}\right) - 0.014\, {\rm ln} \left(-\frac{6 \Lambda_0}{\mu ^2}\right)~,
\end{eqnarray}	
where $\Lambda_0$ has been defined in (\ref{l0def}), $\sigma_c$ denotes the gravitino scalar condensate 
at the non-trivial minimum of the one-loop effective potential (\emph{cf.} fig.~\ref{fig:pot2}),
and 
${\tilde \kappa} = e^{-\langle \varphi  \rangle } \, \kappa$ is the conformally-rescaled gravitational constant in the model of ref.~\cite{confsugra}, defined previously via \eqref{tildcoupl}. In the case of standard ${\mathcal N}=1$ SUGRA, $\langle \varphi \rangle = 0$.  
We note at this stage that the spin-two parts, arising from integrating out graviton quantum fluctuations, are not dominant in the conformal case~\cite{ahm}, provided ${\tilde \kappa}/\kappa \ge {\mathcal O}(10^3)$, which leads~\cite{emdyno} to the agreement of the first  inflationary phase of the model with the Planck data~\cite{Planck}.  
However, if the first phase is succeeded by a Starobinsky phase, it is the latter only that needs to be checked against the data. 

To this end we search numerically for points in the parameter space such that; the effective equations
	\begin{align}
		 \frac{\partial\Gamma}{\partial\Lambda}=0~, \quad
		 \frac{\partial\Gamma}{\partial\sigma}=0~,
	\end{align}
are satisfied, $\Lambda$ is small and positive ($0<\Lambda<10^{-5}M^2_{\rm Pl}$, to ensure the validity of our expansion in $\Lambda$) and $10^{-6}<\mathcal{M}/M_{\rm Pl}<10^{-4}$, to match with known phenomenology of  Starobinsky inflation \cite{Planck}. 
	\begin{figure}[h!!!]
	\centering
		\includegraphics[width=0.48\textwidth]{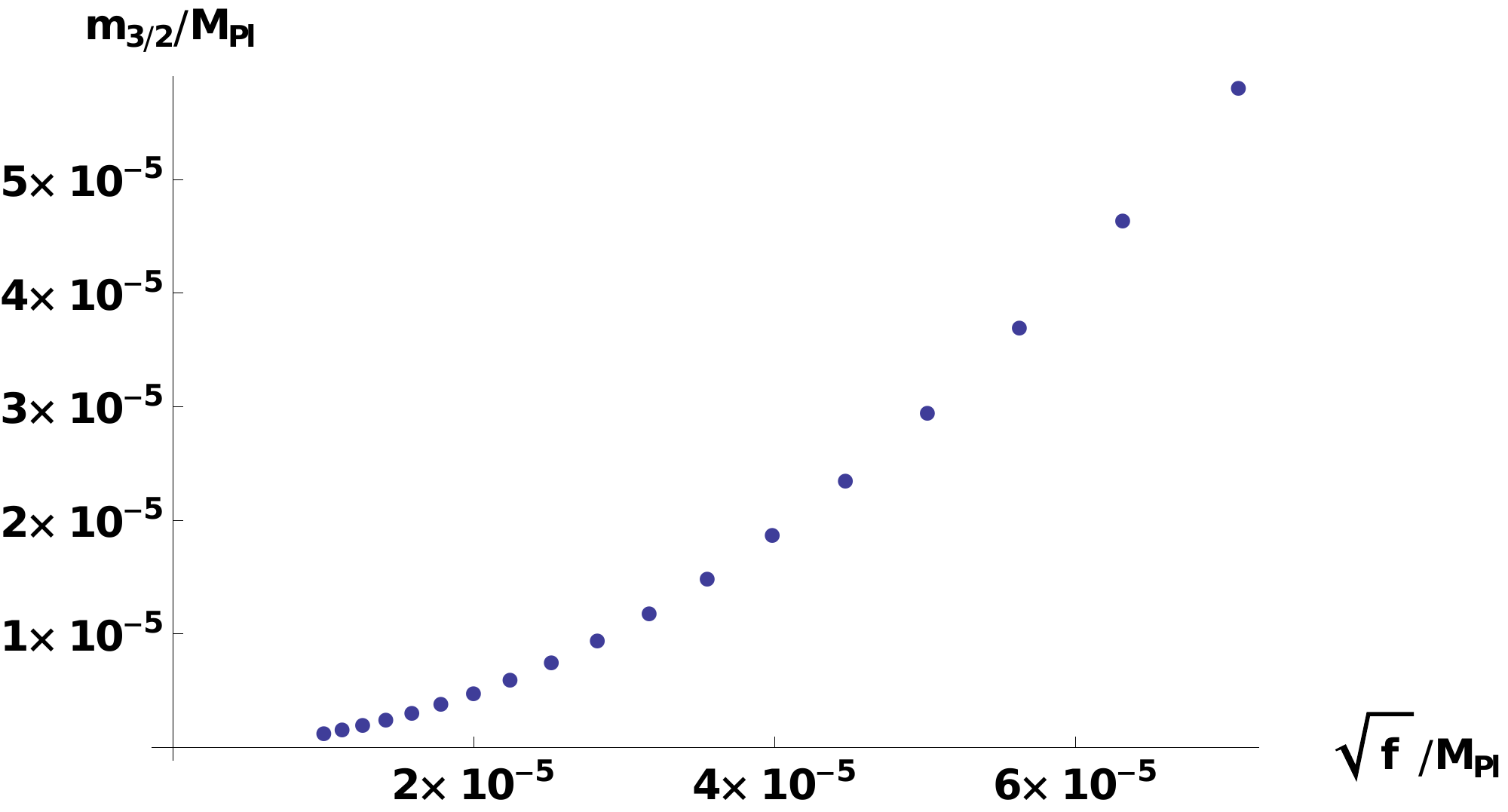} \hfill
		\includegraphics[width=0.48\textwidth]{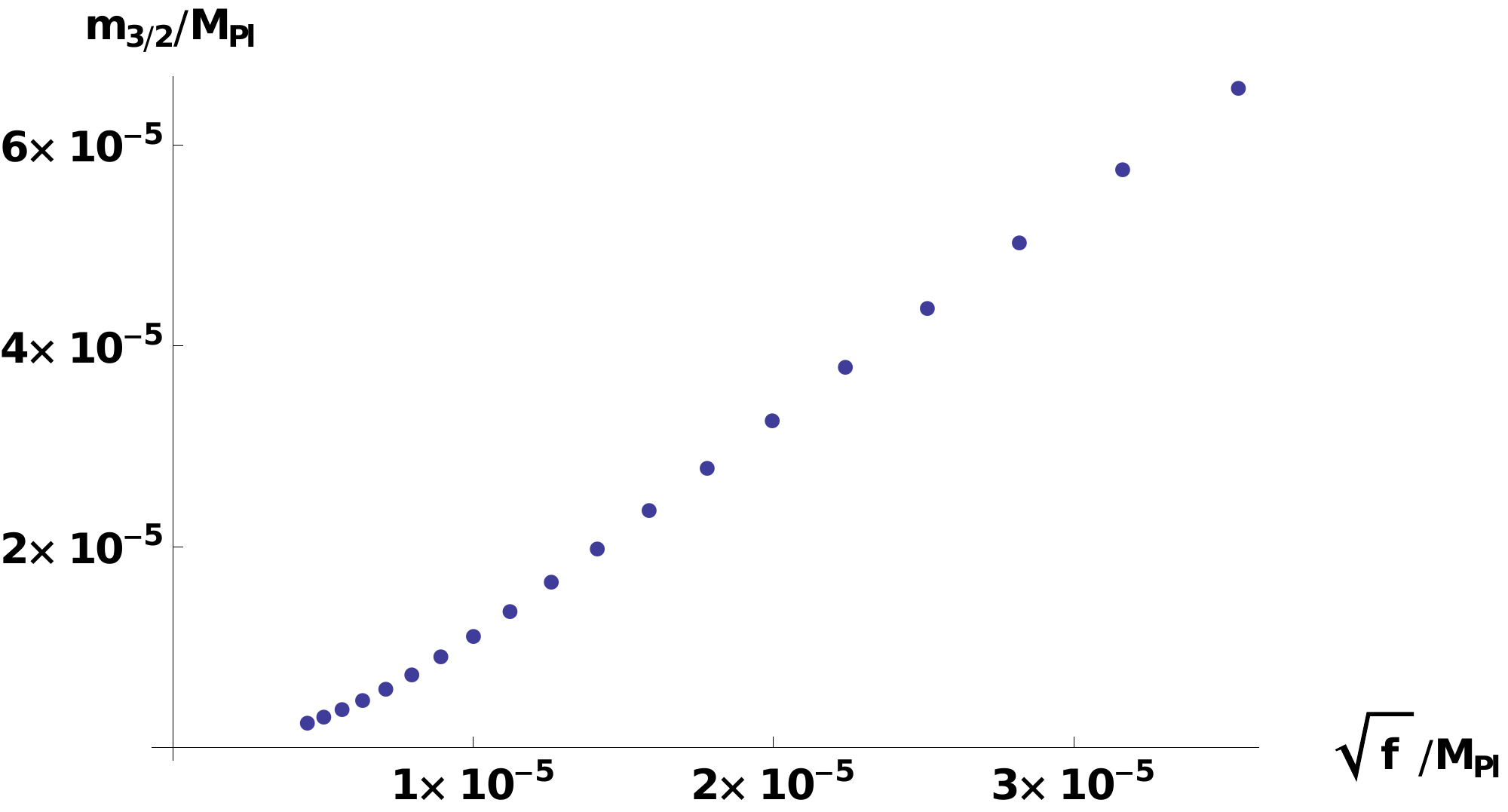}		
		\caption{\emph{Left panel: }Results for $\tilde \kappa=10^3\kappa$. \emph{Right panel:} Results for $\tilde \kappa=10^4\kappa$.}
		\label{fig:2a}
		\end{figure}
For $\tilde \kappa=\kappa$ (\emph{i.e}. for non-conformal supergravity), we were unable to find any solutions satisfying these constraints. 
This of course may not be surprising, given the previously demonstrated non-phenomenological suitability of this simple model~\cite{ahm}. 
If we consider $\tilde \kappa>>\kappa$ however, we find that we are able to satisfy the above constraints for a range of values. 
We present this via the two representative cases below, indicated in fig.~\ref{fig:2a}, where
$\sqrt{f}$ is the scale of global supersymmetry breaking, and we have set the normalisation scale via $\kappa\mu=\sqrt{8\pi}$.\\
Every point in the graphs of the figures is selected to make the Starobinsky scale of order ${\mathcal M} \sim 10^{-5} \, M_{\rm Pl} $, so as to be able to achieve phenomenologically acceptable inflation in the massive gravitino phase, consistent with the Planck-satellite data~\cite{Planck}. 

Exit from the inflationary phase is a complicated issue which we shall not discuss here, aside from the observation that it can be achieved by
coherent oscillations of the gravitino condensate field around its minima, or tunnelling processes \`a  la Vilenkin~\cite{vilenkin}. 
This is still an open issue, which may be addressed via construction of more detailed supersymmetric models, including coupling of the matter sector to gravity. 

\section{Conclusions \label{sec:concl}}

In this talk we considered a minimal inflationary scenario, by means of which a gravitino condensate in supergravity models is held responsible for breaking local supersymmetry dynamically and inducing inflation in an indirect way by means of a Starobinsky-type inflation in the massive gravitino phase. 
Although Inflation of hilltop type via gravitino condensate, where the inflaton is the gravitino condensate field itself, appears at first sight the simplest scenario, nevertheless to ensure slow roll in such a case would require unnaturally high values of the condensate wave-function renormalization, unless the supersymmetry-breaking scale assumed transplanckian values. 
It is in this sense that the Starobinsky-type scenario for inflation, which is associated with the scalar mode that collectively parametrizes the effects of the quadratic-curvature contributions to the effective action of the gravitino condensate, after integrating out graviton and massive gravitino degrees of freedom, appears quite natural. It involves parameters that assume values of a natural and phenomenologically relevant order of magnitude, specifically global supersymmetry scale and gravitino masses of the order of GUT mass scales or less. 

Such a scenario is a truly minimal scenario for natural inflation, in the sense that it involves two scalar primordial composite modes, to achieve dynamical breaking of a gauge symmetry (supergravity) and inflation. From our analysis above, it seems that in order to ensure phenomenologically relevant (for the stability of the electroweak vacuum) supersymmetry breaking scales and gravitino masses, one needs to apply the above ideas, not to the minimal SUGRA model, but to Jordan-frame extensions thereof, involving a third scalar field (dilaton). In the context of the next to minimal supersymmetric standard model, which Jordan-frame SUGRA models can incorporate, such dilatons may be composite of appropriate matter superfields, involving Higgs (supermultiplets). 

Details of the microscopic matter model are important to ensure the correct cosmological evolution, in particular satisfaction of the Big-Bang-Nucleosynthesis constraints. A GUT scale gravitino can be made to decay fast enough so as not to disturb the BBN, but this depends on the details of the matter sector of the theory, which we have not discussed so far. We plan to do so in the future. 

Nevertheless, we believe that the above dynamical breaking of supergravity scenario and the links with Starobinsky inflation are interesting paradigms, which have a chance of leading to realistic phenomenological scenarios compatible with the cosmological and particle physics data.

\section*{Acknowledgements}

N.E.M. would like to thank the organisers of the third International Conference of New frontiers in Physics (ICNFP 2014) for their invitation to a plenary talk.
This work is supported in part by the London Centre for Terauniverse Studies (LCTS), using funding from the European Research Council via the Advanced Investigator Grant 267352 and by STFC (UK) under the research grant ST/J002798/1.

  \end{document}